\documentclass[preprint, 12pt]{elsarticle}
\journal{Journal of Molecular Spectroscopy}
\usepackage{amsmath}
\usepackage{siunitx}
\usepackage{url}
\begin{document}
\begin{frontmatter}
\title{Microwave spectroscopy of $\Lambda$-doublet transitions in the ground state of CH}
\author{S. Truppe}
\author{R. J. Hendricks}
\author{S. K. Tokunaga}
\author{E. A. Hinds}
\author{M. R. Tarbutt}
\ead{m.tarbutt@imperial.ac.uk}
\address{Centre for Cold Matter, Blackett Laboratory, Imperial College London, Prince Consort Road, London, SW7 2AZ, United Kingdom.}
\date{}
\begin{abstract}
The $\Lambda$-doublet transitions in CH at 3.3 and \SI{0.7}{\giga\hertz} are unusually sensitive to variations in the fine-structure constant and the electron-to-proton mass ratio. We describe methods used to measure the frequencies of these transitions with Hz-level accuracy. We produce a pulsed supersonic beam of cold CH by photodissociation of CHBr$_3$, and we measure the microwave transition frequencies as the molecules propagate through a parallel-plate transmission line resonator. We use the molecules to map out the amplitude and phase of the standing wave field inside the transmission line. We investigate velocity-dependent frequency shifts, showing that they can be strongly suppressed through careful timing of the microwave pulses. We measure the Zeeman and Stark effects of the microwave transitions, and reduce systematic shifts due to magnetic and electric fields to below \SI{1}{\hertz}. We also investigate other sources of systematic uncertainty in the experiment.
\end{abstract}
\end{frontmatter}

\section{Introduction}
Many extensions to the standard model of particle physics make the intriguing prediction that quantities we normally consider to be fundamental constants - such as the fine-structure constant $\alpha$ or the electron-to-proton mass ratio $\mu$ - may in fact vary with time, position or with the local density of matter \cite{Uzan2003}. These theories aim to unify gravity with the other forces, or to explain the nature of dark energy. The lowest $\Lambda$-doublet and millimetre-wave transitions in the CH molecule are particularly sensitive to variation of $\alpha$ and $\mu$ \cite{Kozlov2009, deNijs2012}. Recently, we tested the hypothesis that these constants may differ between the high density environment of the Earth and the vastly lower density of the interstellar medium, by comparing microwave frequencies of CH observed in cold interstellar gas clouds in our own galaxy to those measured in the laboratory \cite{Truppe2013}. Using this method, we were able to constrain variations in these fundamental constants at the 0.1 parts-per-million level.

To measure the laboratory frequencies with Hz-level accuracy, we developed a source which produces short pulses of CH molecules at low temperature, and we developed a variation on the method of Ramsey spectroscopy \cite{Ramsey1950}. In our method, the pulse of molecules propagate along a parallel-plate transmission line which supports a standing wave of the microwave field. In this geometry, the amplitude, polarization, and phase of the field are exceptionally well controlled. The microwaves can be pulsed on when the molecules are at any chosen position, and because the molecular pulse is short and the velocity spread is low, they can be used to map out the amplitude and phase of the field as a function of position \cite{Hudson2007}. Doppler shifts are cancelled because the field is (nearly) a standing wave, and any residual velocity-dependent shifts are easily measured by changing the beam velocity. Here, we describe the source of CH, the spectroscopic technique, and the methods we use to control systematic frequency shifts.
\section{The CH molecule}
The methylidyne radical, CH, plays an essential role in both chemistry and physics. Early investigations of optical emission spectra of CH helped to explain the spectra of diatomic molecules (see \cite{Herzberg1969} and references therein). CH is a major participant in most combustion processes. It was one of the first molecules to be detected in the interstellar medium, in stellar atmospheres, and in comets, by means of optical absorption spectroscopy \cite{Herzberg1950, Brown2003}. It is an essential building block in the formation of complex carbon-chain molecules in interstellar gas clouds \cite{Gerlich1992} and it is commonly used as a tracer for atomic carbon and molecular hydrogen \cite{Chastain2010}. In 1937 Dunham and Adams recorded the first optical spectrum of CH in the interstellar medium \cite{Dunham1937}. Over forty years later the frequencies of the two lowest-lying $\Lambda$-doublet transitions, at \SI{3.3}{\giga\hertz} and \SI{0.7}{\giga\hertz}, were determined by radio-astronomy \cite{Rydbeck1973, Turner1974, Ziurys1985}. Laboratory measurements followed \cite{Brazier1983, McCarthy2006}, but the radio-astronomy measurements remained the most precise until our own recent measurements \cite{Truppe2013} using the methods described here.

\begin{figure}%
\centering
\includegraphics{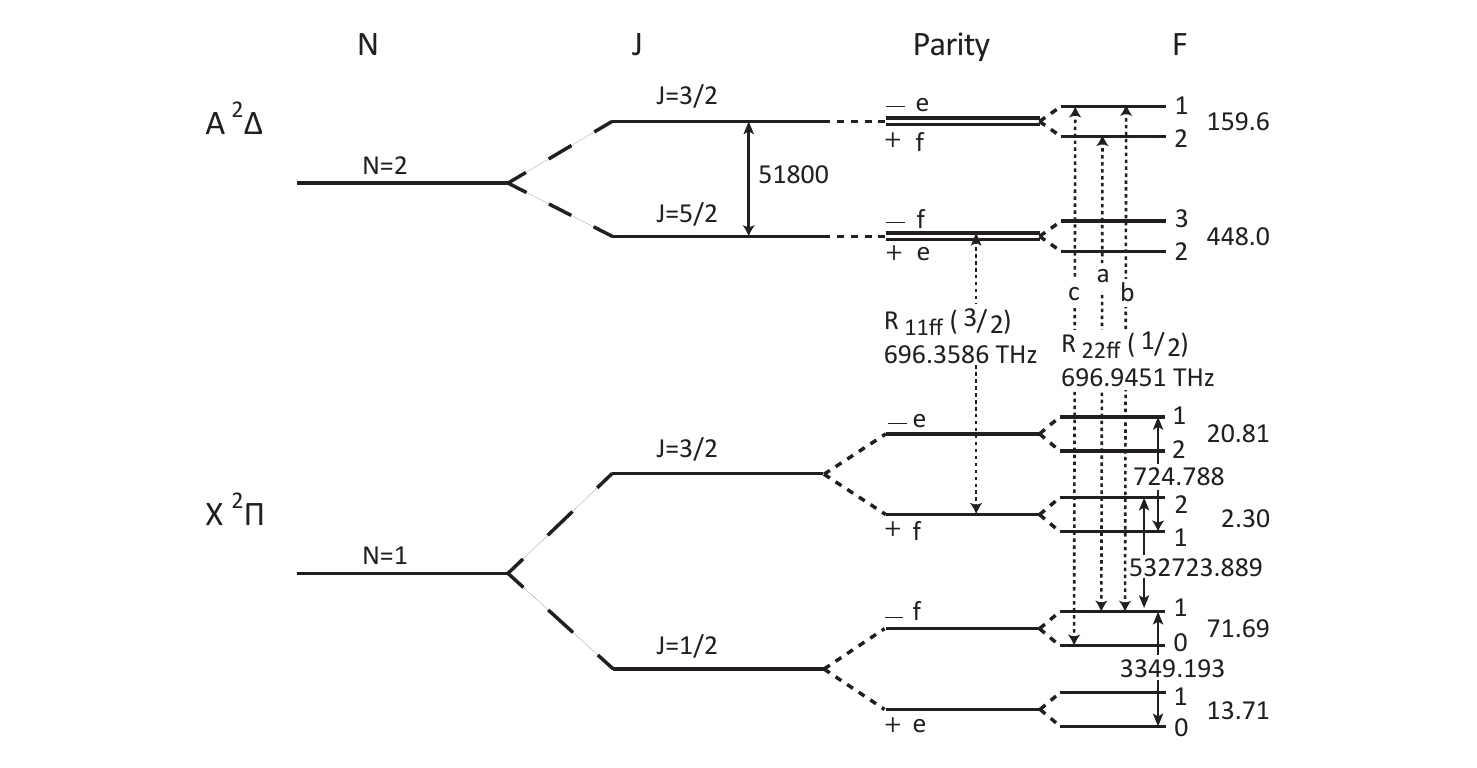}%
\caption{Relevant energy levels and transitions between the X and A states of CH. Each state labelled by $J$ is split into two $\Lambda$-doublet levels of opposite parity ($\pm$, also labelled by their e/f parity). Each of these is further split by the hyperfine interaction into a pair of states with total angular momentum quantum numbers $F=J\pm1/2$. Dotted lines show the optical transitions used for detecting molecules in the $J=3/2$ and $J=1/2$ states, labelled R$_{11ff}(3/2)$ and R$_{22ff}(1/2)$ respectively. The hyperfine components of the latter transition are labelled as a, b and c. Transition frequencies are given in MHz unless stated otherwise.}%
\label{chlevel}%
\end{figure}

Figure~\ref{chlevel} shows the low-lying energy level structure in the ground electronic state of CH, $\text{X} ^2\Pi (v=0)$, and the first electronically excited state, $\text{A} ^2\Delta (v=0)$, both described using the Hund's case (b) coupling scheme. Here, the electronic orbital angular momentum $\mathbf{L}$ and the rotational angular momentum $\mathbf{R}$ are coupled to a resultant $\mathbf{N}=\mathbf{L}+\mathbf{R}$. The spin angular momentum $\mathbf{S}$ adds to $\mathbf{N}$ to give the total electronic angular momentum $\mathbf{J}=\mathbf{N}+\mathbf{S}$. Each $J$-level is split into a $\Lambda$-doublet which is composed of two closely-spaced states of opposite parity $\left|p=\pm 1\right\rangle=\left|+|\Lambda|\right\rangle\pm (-1)^{J-S}\left|-|\Lambda|\right\rangle$, where $\Lambda$ is the quantum number for the projection of $\mathbf{L}$ onto the internuclear axis. The interaction with the $I=1/2$ hydrogen nuclear spin splits each $\Lambda$-doublet component into a pair of hyperfine levels, labelled by the total angular momentum quantum number $F=J\pm 1/2$. We use the short-hand notation $(J^p,F)$ to label the energy levels of the X state.

\section{Production and detection of CH}

\begin{figure}%
\centering%
\includegraphics[width=\columnwidth]{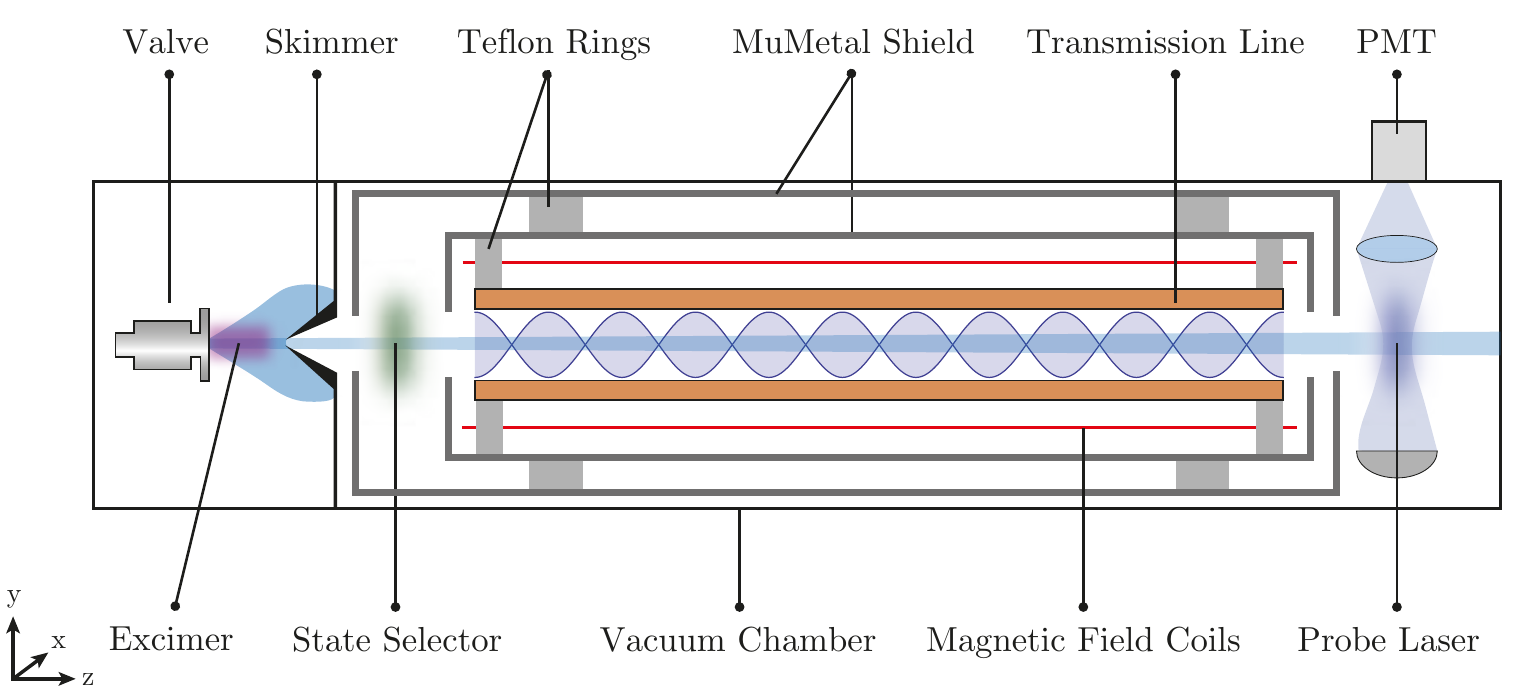}%
\caption{Sketch of the experiment to measure the $\Lambda$-doublet transitions of CH (not to scale). The molecular beam is produced via photodissociation of bromoform. The beam passes through a skimmer and a state selector, and then travels between the plates of a magnetically-shielded parallel-plate transmission line where the microwave transition is driven. Finally, the molecules are detected by laser induced fluorescence. For the measurement of the $J=1/2$ $\Lambda$-doublet transition the outer magnetic shield and the inner magnetic field coils were absent.}%
\label{ms32}%
\end{figure}

Figure~\ref{ms32} shows the apparatus. We produce a pulsed, supersonic beam of CH molecules by photo-dissociation of bromoform (CHBr$_3$), following some of the methods described in \cite{Lindner1998, Romanzin2006}. A carrier gas, with a backing pressure of \SI{4}{bar}, is bubbled through liquid bromoform (Sigma Aldrich, 96\% purity, stabilized in ethanol) which is held at room temperature in a stainless steel container. The mixture expands through the \SI{1}{\milli\meter} orifice of a solenoid valve (General Valve Series 99) into a vacuum chamber at a repetition rate of \SI{10}{\hertz}. To dissociate the bromoform, we use pulses of light from an excimer laser, with an energy of \SI{220}{\milli\joule}, a wavelength of \SI{248}{\nano\meter}, and a duration of \SI{20}{\nano\second}. This light propagates along the $x$-axis, and is focussed to a rectangular spot, \SI{1}{\milli\meter} high (along $y$) and \SI{4}{\milli\meter} wide (along $z$), in the region immediately beyond the nozzle of the valve. The excimer pulse sets our origin of time. Further details of the source are given in \cite{TruppeThesis}.

The molecules pass through a skimmer with a \SI{2}{\milli\meter} diameter orifice, situated at $z=\SI{78}{\milli\meter}$, then through the apparatus used to measure the microwave transition frequencies (described below), and are finally detected at $z=D=\SI{780}{\milli\meter}$ by laser-induced fluorescence. The probe laser used for detection is a frequency-doubled continuous-wave titanium-sapphire laser, tuned to the $A^2\Delta(v=0)\leftarrow X^2\Pi(v=0)$ transition near~\SI{430.15}{\nano\meter}. This probe is linearly polarized along $z$, propagates along $x$, has a power of \SI{5}{\milli\watt}, and is shaped to a rectangular cross-section, \SI{4}{\milli\meter} in the $y$-direction and \SI{1.4}{\milli\meter} in the $z$ direction. The induced fluorescence is imaged onto a photomultiplier tube, and the signal is recorded with a time resolution of approximately \SI{5}{\micro\second}.

\begin{figure}%
\centering%
\includegraphics[width=\columnwidth]{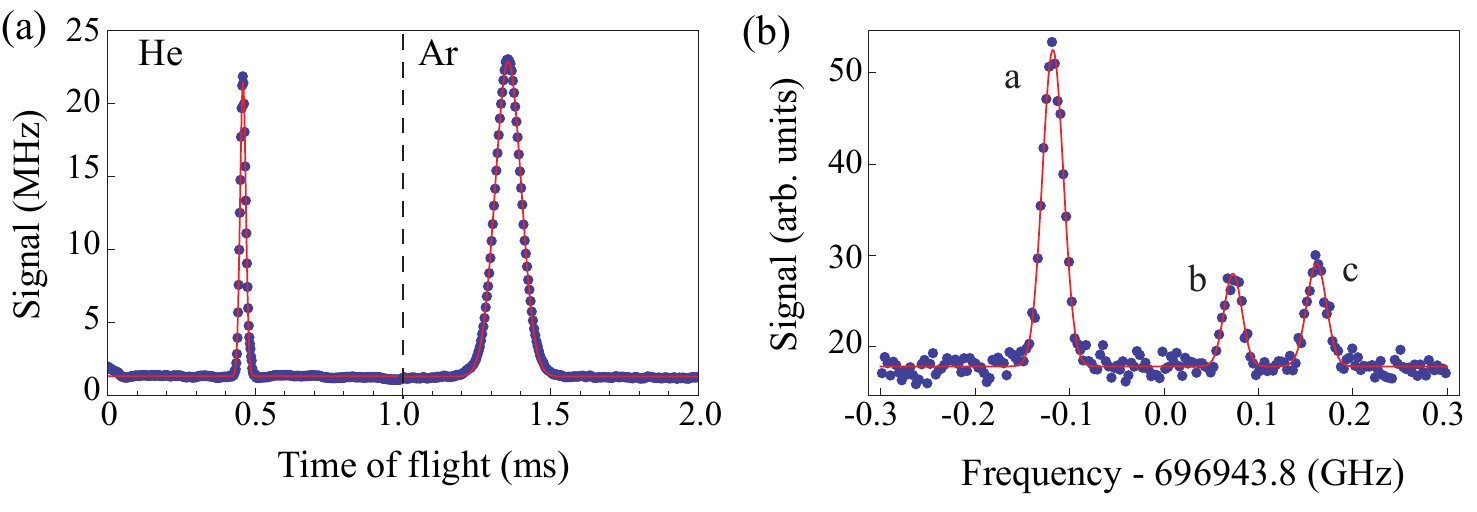}%
\caption{(a) Time of flight profiles of CH molecules using (separately) He and Ar carrier gases. Points: data, line: Gaussian fits. (b) Spectrum showing the three hyperfine components of the R$_{22ff}(1/2)$ line of the $A^2\Delta (v=0,N=2)\leftarrow X^2\Pi(v=0, N=1)$ transition. The labels a, b, c correspond to those in figure \ref{chlevel}. Points: data. Line: Fit to a sum of three Gaussians.}%
\label{tofSpec}%
\end{figure}

To vary the beam velocity, $v_0$, we use He, Ne, Ar and Kr carrier gases. Figure \ref{tofSpec}(a) shows the time-of-flight profiles of CH molecules arriving at the detector when He and Ar are used. From the arrival times we measure the mean speeds to be $v_0=1710$, 800, 570 and~\SI{420}{\meter\per\second} for He, Ne, Ar and Kr respectively. The duration of the pulse of CH produced in the source is determined by the \SI{4}{\milli\meter} width of the excimer beam and is always very short compared with the width of the time-of-flight profile measured at the detector. This width therefore measures the translational temperature of the CH beam. From the data in figure \ref{tofSpec}(a) we measure a CH translational temperature of \SI{2}{\kelvin} and \SI{0.4}{\kelvin} for He and Ar carrier gases respectively.

We selectively detect molecules in either the $(1/2^{-},F)$ or the $(3/2^{+},F)$ levels, by driving one of the two transitions labelled R$_{22ff}(1/2)$ and R$_{11ff}(3/2)$ in figure \ref{chlevel}. Figure \ref{tofSpec}(b) shows the spectrum recorded as the probe laser is scanned over the three well resolved hyperfine components of the R$_{22ff}(1/2)$ transition. Because the detected molecules have a range of transverse speeds, the spectral lines are Doppler broadened to a full width at half maximum of \SI{23}{\mega\hertz}. For the microwave spectroscopy, the laser frequency is locked to one of the hyperfine components of the relevant transition using an optical cavity which is itself locked to a stabilized He-Ne laser.

\section{Microwave spectroscopy setup}

At $z=\SI{241}{\milli\meter}$, the molecules pass through the state-selector, which selectively populates one of the two parity eigenstates of a $\Lambda$-doublet. For the $J=1/2$ measurements, the state selection is done by optically pumping molecules out of the relevant $(1/2^-,F)$ state, using \SI{40}{\milli\watt} of laser light at the same frequency as the probe. This light is reflected almost back on itself to increase the interaction time with the molecules. The optical pumping removes about 70\% of the initial population. Over 95\% of the CH molecules produced in our source are in the ground $J=1/2$ state, so for the $J=3/2$ measurements the state selection is done by driving population into the $(3/2^+,1)$ or $(3/2^+,2)$ states using \SI{10}{\micro\watt} of radiation near \SI{533}{\giga\hertz}. This millimetre-wave radiation is generated by an amplifier-multiplier chain that produces the 54th harmonic of a frequency synthesizer. The transfer efficiency is about 40\%. A description of the measurement of this lowest millimeter-wave transition is given in reference \cite{Truppe2013a}.

\begin{figure}%
\centering%
\includegraphics{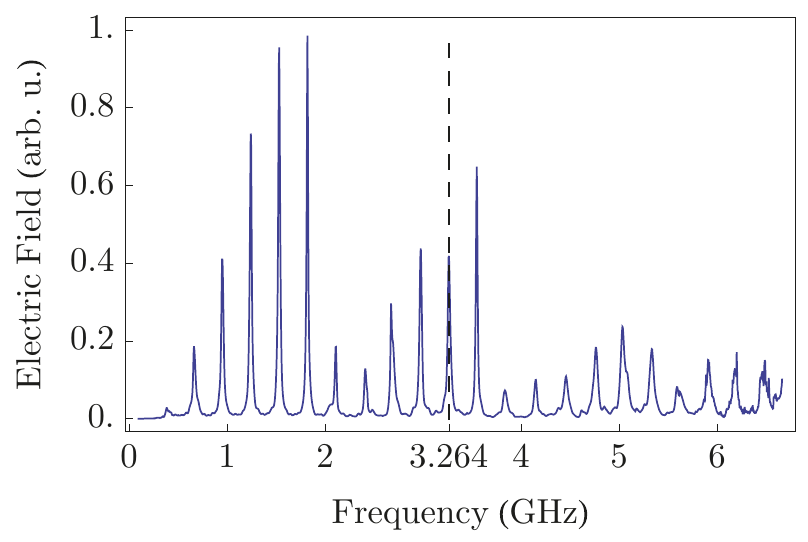}%
\caption{Spectrum of the transmission line resonator. The plate length is chosen so that the resonator has a peak near the resonance frequency of the molecules. The width of the transmission line resonance is about~\SI{35}{\mega\hertz}. The dashed line indicates the approximate transition frequency of the $(1/2^+,1)-(1/2^-,0)$ line at \SI{3.264}{\giga\hertz}.}%
\label{stand}%
\end{figure}

Following the state-selector the molecules enter the transmission line resonator where we drive the $\Lambda$-doublet transition. The transmission line is formed by a pair of parallel copper plates of length $L$ and width $w$, separated vertically by a distance $d$. It is fed from a semi-rigid, non-magnetic coaxial cable, with the inner conductor connected to one plate and the outer conductor connected to the other. The other end of the transmission line is open, and the wave reflects from this end to form a standing wave. The quality factor of this resonator is determined by the reflectivity of the open end and the transmission between the coaxial cable and the transmission line. We cut the plates to a length such that a resonance frequency of the transmission line matches the approximate $\Lambda$-doublet transition frequency. This length is approximately $L=n/(2 \lambda)$ where $n$ is the number of electric field antinodes in the resonator, but this is insufficiently accurate because the position where the wave reflects from the ends is not well defined. Instead, we measure the spectrum of the resonator with a vector network analyzer and then reduce the length to obtain the desired resonance frequency. The lengths were approximately \SI{480}{\milli\meter} for the $J=1/2$ measurements and \SI{440}{\milli\meter} for the $J=3/2$ measurements. Figure \ref{stand} shows the spectrum of the resonator set up for measuring the $(1/2^+,1)-(1/2^-,0)$ transition. The free spectral range of the resonator is \SI{300}{\mega\hertz}, and the full width at half maximum (FWHM) of the peaks is \SI{35}{\mega\hertz}, corresponding to a round-trip loss of 50\%.

We aim to propagate only the TEM$_{00}$ mode so that the electric field is uniform between the plates and accurately polarized along $y$. If the plate spacing is large enough, the transmission line can support the higher-order TE and TM modes that have components of the k-vector along $y$, but these modes are cut off if the plate spacing is smaller than half the wavelength, $d < \lambda / 2$. The shortest wavelength we use in this experiment is \SI{90}{\milli\meter}, whereas the plate spacing is only \SI{5}{\milli\meter}, so these higher-order modes are very strongly attenuated. There can also be modes that have components of the k-vector along $x$. These lesser-known modes zig-zag horizontally in the space between the two plates, reflecting from the open edges of the structure. Because they can radiate into the open space beyond the plates, they are referred to as leaky modes. The procedure for calculating the k-vectors of these modes is described in \cite{Rushdi1978}. In an early version of the experiment, we used plates of width $w=\SI{50}{\milli\meter}$ spaced by $d=\SI{10}{\milli\meter}$. In this case, we find theoretically that there is a leaky mode which has a propagation wavelength of \SI{130}{\milli\meter} and a $1/e$ attenuation distance of \SI{220}{\milli\meter}. Using the characterization methods described below, we indeed observed a mode with this wavelength. To cut-off the leaky mode, we reduced the width of the plates to $w=\SI{30}{\milli\meter}$ and also reduced the plate spacing to $d=\SI{5}{\milli\meter}$. This increases the leaky mode propagation wavelength to \SI{600}{\milli\meter} and reduces the attenuation distance to just \SI{20}{\milli\meter}. Since the latter is far smaller than the length of the interaction region, the mode is very effectively eliminated from the experiment. The spectrum shown in figure \ref{stand} shows that there are no significant higher-order modes.

The microwave radiation is generated by a frequency synthesizer which is phase locked to a GPS-disciplined frequency reference to a fractional uncertainty better than $10^{-13}$. The synthesizer is connected to the transmission line resonator through a fast, high isolation switch.

\section{Characterizing the microwave field}

The electric field inside the transmission line resonator can be described by
\begin{equation}
E=\frac{A}{2} (1+\Delta)\cos(k (z-z_0) -\omega t)+\frac{A}{2}(1-\Delta)\cos(k(z-z_0) +\omega t)\, ,
\label{estan}
\end{equation}
where $k=2\pi/\lambda$ is the wave vector, $A$ is the amplitude of the electric field, $z_{0}$ is the position of an antinode of the standing wave, and $\Delta$ is an amplitude imbalance to account for the fact that the wave is not perfectly reflected at the end of the transmission line. This equation can be rewritten as
\begin{equation}
E=A\sqrt{\cos^2\left(k (z-z_0)\right) + \Delta^2\sin^2\left(k (z-z_0)\right)}\cos(\omega t-\phi(z))\, ,
\label{field}
\end{equation}
where
\begin{equation}
\phi(z)=\tan^{-1}\left[\Delta\tan\left(k (z-z_0)\right)\right] +\phi_{0}\,.
\label{imbal}
\end{equation}
Here, $\phi_{0}=0$ when $\cos(k (z-z_0))>0$ and $\pi$ otherwise.

For a single microwave pulse, the interaction of the molecules with the radiation transfers population from the initial to the final state with a probability of
\begin{equation}
P_{\text{1-pulse}}=\frac{\Omega^2}{\Omega^2+\delta^2}\sin^2\left(\frac{\sqrt{\Omega^2+\delta^2}\tau}{2}\right)\, ,
\label{prob}
\end{equation}
where, $\Omega=d_{12} E/\hbar$ is the Rabi frequency, $d_{12}$ is the transition dipole moment, $\delta=\omega-\omega_0$ is the detuning of the microwave angular frequency $\omega$ from the resonance angular frequency $\omega_0$, and $\tau$ is the interaction time. When $\Omega\tau=\pi$ and $\delta=0$ ($\pi$-pulse condition) the entire population is transferred from the initial to the final state.

\begin{figure}%
\centering%
\includegraphics[width=\columnwidth]{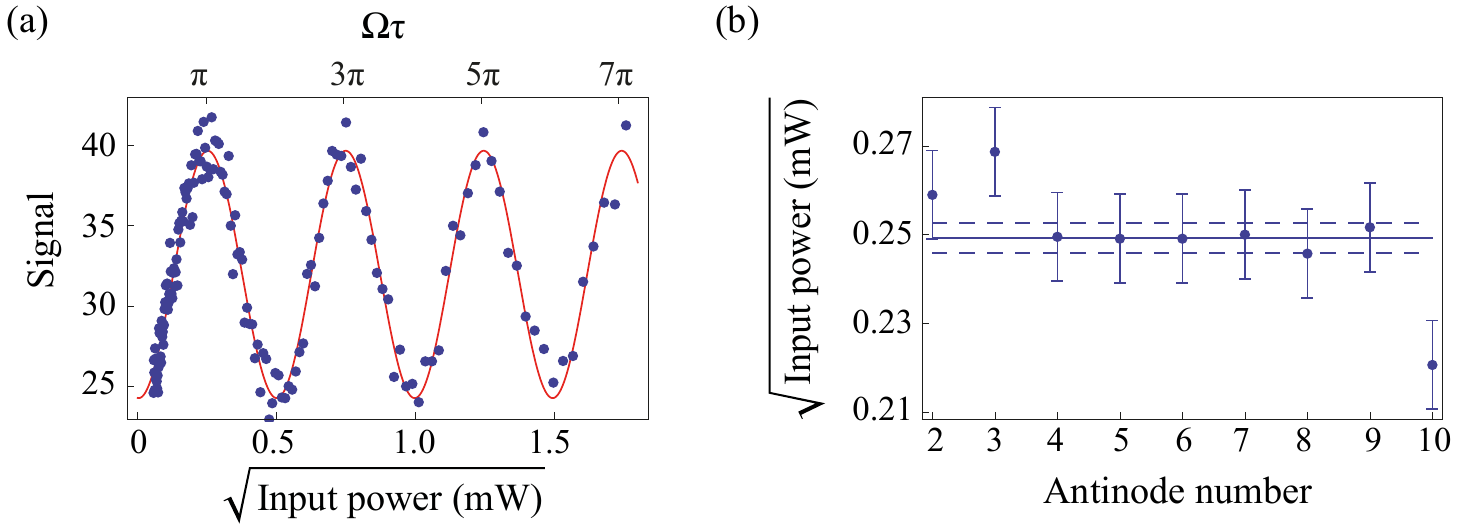}%
\caption{(a) Rabi oscillations. The detuning is set to $\delta=0$, and the interaction time to $\tau=\SI{15}{\micro\second}$. By scanning the microwave power we observe oscillations in the $(1/2^-,1)$ population. This allows us to find the right power for a $\pi$-pulse. We fit $S=S_0+A\sin^2(\alpha x)$ (red solid line) to the data (blue dots), where $x$ is proportional to the square root of the applied microwave power, and $S_0$, $A$ and $\alpha$ are fitting parameters. (b) Power needed to drive a $\pi$-pulse for each antinode of the standing wave. Solid and dashed lines are the mean and standard deviation of the set.}%
\label{rabi}%
\end{figure}

To begin an experiment, we first pulse the microwaves on for just a short period, $\tau=\SI{15}{\micro\second}$, at the time when the molecular pulse is at the centre of the transmission line, near an antinode of the electric field. Scanning the microwave frequency over the resulting broad resonance gives a first estimate of the transition frequency. The frequency is then fixed at the transition frequency and the microwave power is scanned. We observe Rabi oscillations in the measured population, an example of which is shown in figure~\ref{rabi}(a). Since $\Omega$ is proportional to the electric field of the radiation, we plot the signal versus the square root of the microwave power, and then fit equation (\ref{prob}) to the data. This identifies the exact power needed to drive a $\pi$-pulse for this pulse duration and for this particular position of the molecules. Figure \ref{rabi}(b) shows the power needed for a $\pi$-pulse with the molecules centred on each of the antinodes. The variation along the the transmission line is small.

\begin{figure}%
\centering%
\includegraphics[width=\columnwidth]{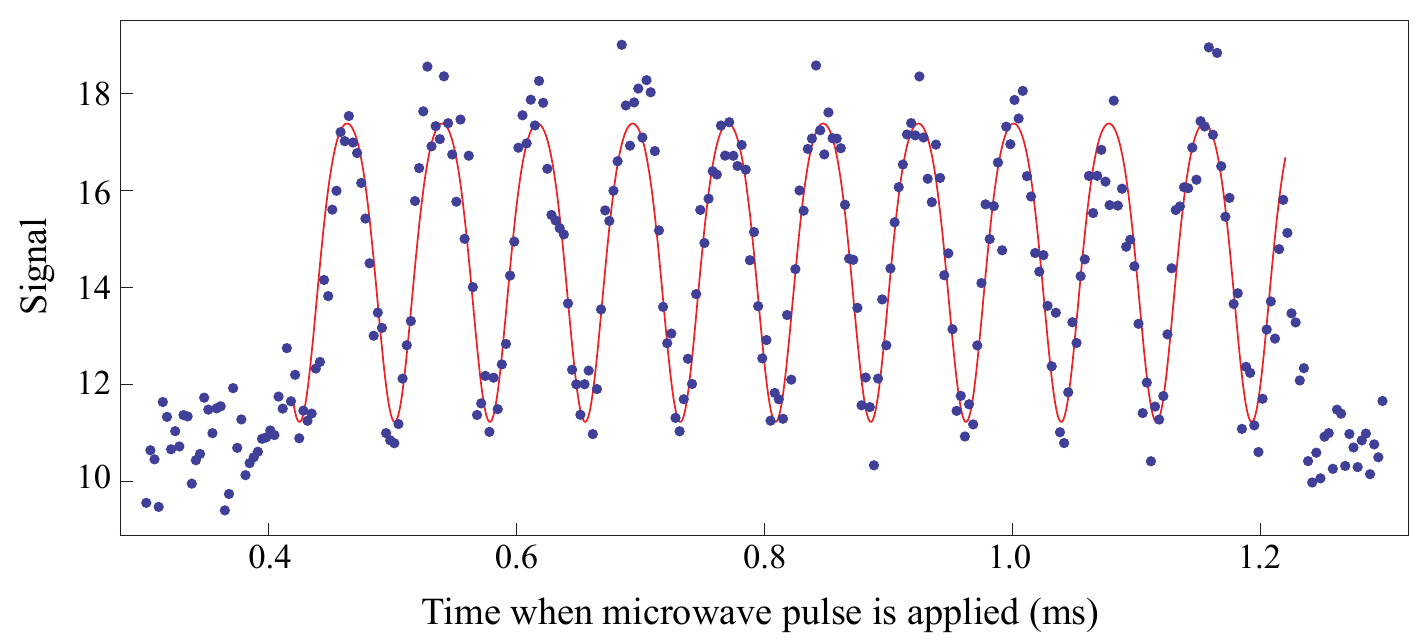}%
\caption{Mapping the amplitude of the microwave field by recording the fluorescence as a function of the time when a resonant \SI{15}{\micro\second} microwave pulse is applied. The power corresponds to a $\pi$-pulse when the molecules are at the central antinode. Blue dots: data. Red line: fit using the model of equation~(\ref{standingSig}).}%
\label{fieldmap}%
\end{figure}

Next, we use the molecules to make a map of the microwave electric field amplitude inside the transmission line. We do this by measuring the population as a function of the time when a short, resonant microwave pulse is applied. Figure \ref{fieldmap} shows an example of the data obtained. Here, we have set $\delta=0$, $\tau=\SI{15}{\micro\second}$, and the microwave power such that $\Omega \tau = \pi$ when the molecules are at the central antinode. When $\delta=0$ the transition probability becomes $P=\sin^2\left(\Omega\tau/2\right)$. The Rabi frequency $\Omega$ is proportional to the electric field which varies along the transmission line according to equation (\ref{estan}). For this measurement, we take $\Delta=0$ and so $\Omega=\Omega_{\text{max}}\cos\left[2\pi\left(z-z_0\right)/\lambda\right]$. To account for the finite spread of the molecules we introduce an averaged Rabi frequency such that $\Omega_{\text{max}}\tau=q\pi$ with $q<1$. The line in figure \ref{fieldmap} is a fit to the data using the expected model
\begin{equation}
S=S_0+A\sin^2\left(q\frac{\pi}{2}\cos\left(\frac{2\pi\left(v t-v t_0\right)}{\lambda}\right)\right)\, ,
\label{standingSig}
\end{equation}
where $t=z/v$ and $v=\SI{570}{\meter\per\second}$. The parameter $q$, the offset $S_0$, the amplitude $A$, the initial time $t_0$, and the wavelength $\lambda$ are all fitting parameters. The fit yields $\lambda=8.99\pm0.01\SI{}{\centi\meter}$, in agreement with the expected wavelength at \SI{3.335}{\giga\hertz}. This field map is essential for the Ramsey experiments described below, where we need to know exactly when the molecular pulse passes each antinode.

\section{Frequency measurements}

\begin{figure}%
\centering%
\includegraphics{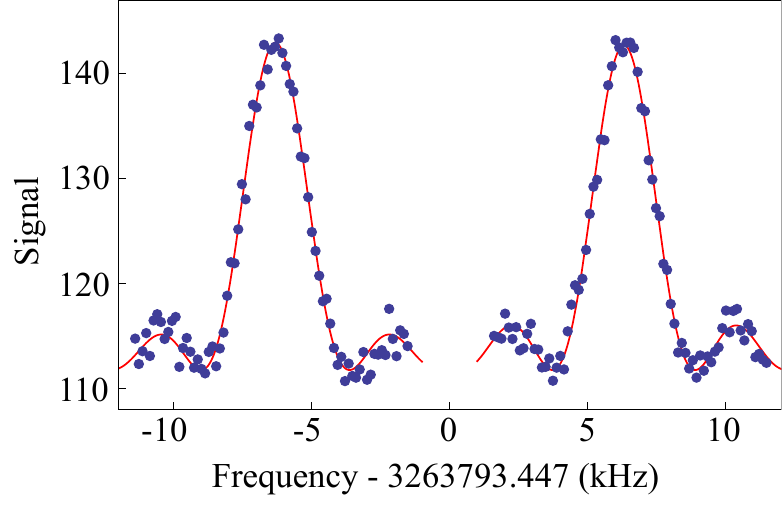}%
\caption{Single pulse measurement of the $(1/2^+,1)-(1/2^-,0)$ transition with $\tau=326$~\SI{}{\micro\second}. The red line is a fit using the model discussed in the text.}%
\label{sPlot}%
\end{figure}

Next, we increase the interaction time, $\tau$, and decrease the microwave power accordingly to maintain $\Omega \tau = \pi$. Figure \ref{sPlot} shows the signal as a function of microwave frequency for the $(1/2^+,1)-(1/2^-,0)$ transition, when $\tau=\SI{326}{\micro\second}$. The timing is chosen so that the molecules pass through the central antinode half way through this interaction time. Because the field consists of two counter-propagating waves, parallel and anti-parallel to the molecular beam direction, we see two resonances separated by twice the Doppler shift $\Delta\omega_D=\pm\omega_{0} \frac{v}{c}$, where $v$ is the velocity of the molecules. To each resonance we fit the function $S=S_0+\frac{\Omega^2}{\Omega^{2}+\delta^{2}}\sin^2\left(\frac{\sqrt{\Omega^{2}+\delta^{2}}\tau}{2}\right)$, with $\tau=~\SI{326}{\micro\second}$, $\Omega\tau=q\pi$, and $\delta=\omega-\omega_0$. Here, $S_0$, $\omega_0$ and $q$ are fitting parameters. The amplitudes of the two peaks are equal within their uncertainties of 3\%. We find that the power needed for a $\pi$-pulse is the same for the two peaks to within 2\% showing that $\Delta$ is less than 0.005 at this frequency. The mean of the two centre frequencies gives the Doppler-free resonance frequency, which we find to be $3263793456 \pm \SI{17}{\hertz}$ for the $(1/2^+,1)-(1/2^-,0)$ transition and $3335479349\pm\SI{7}{\hertz}$ for the $(1/2^+,1)-(1/2^-,1)$ transition. With transitions produced by a single pulse of radiation, it is well known that splitting the line to such high accuracy may suffer from systematic errors. Specifically, inhomogeneities in either the static field or the ac field can produce an asymmetric lineshape and a corresponding shift of the line centre. Remarkably, however, these resonance frequencies agree at the level of 7\,Hz with the measurements below, using the Ramsey method. This agreement indicates that there is no such lineshape distortion at this level.

\begin{figure}%
\centering%
\includegraphics[width=\columnwidth]{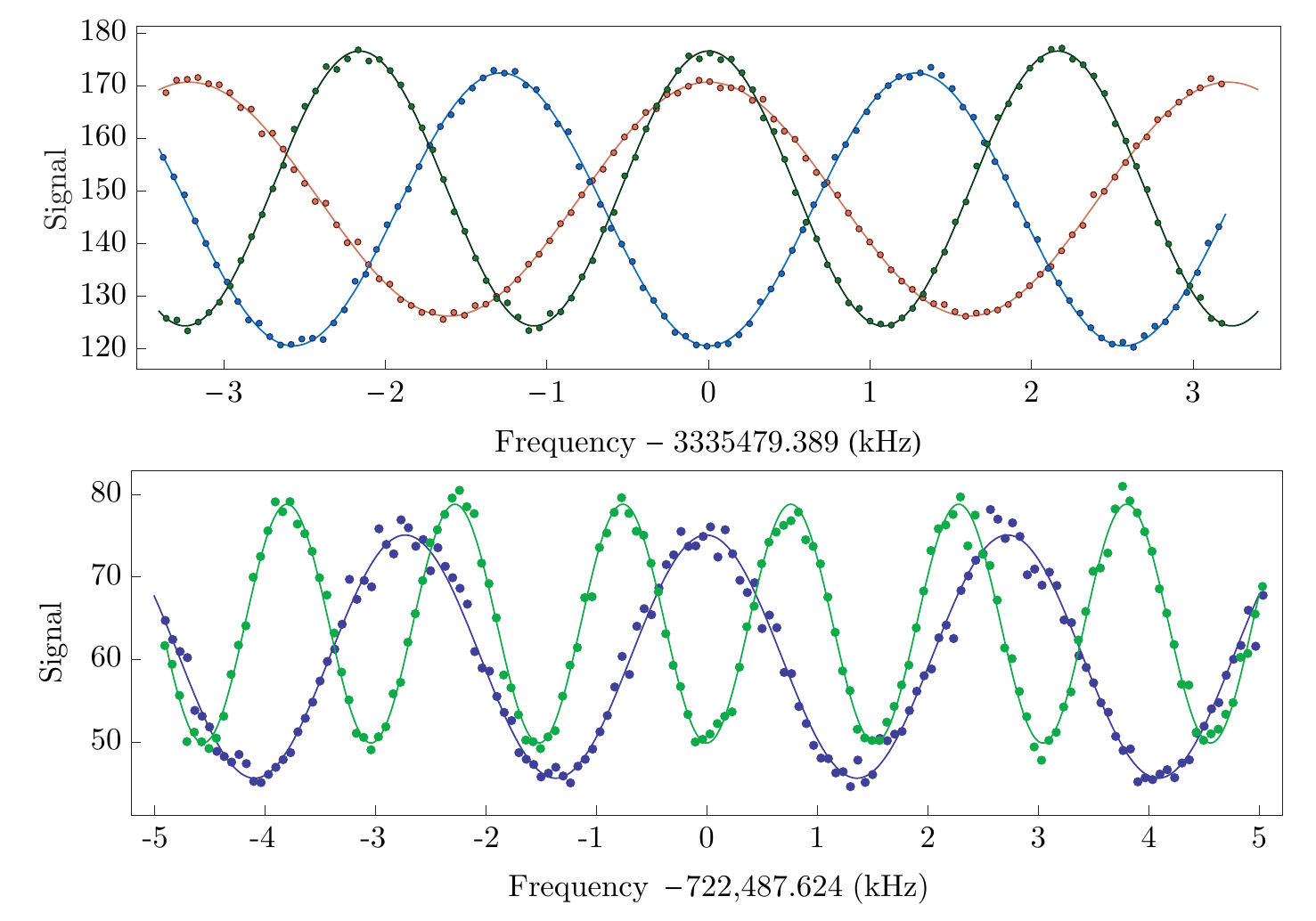}%
\caption{Frequency measurements using the method of separated oscillatory fields. Top: Population in the $(1/2^-,1)$ state as a function of the microwave frequency for three different free evolution times, \SI{458}{\micro\second} (green), \SI{380}{\micro\second} (blue), \SI{302}{\micro\second} (red). Bottom: Population in the $(3/2^+,1)$ state as a function of the microwave frequency for two different free evolution times, \SI{650}{\micro\second} (green), \SI{330}{\micro\second} (blue). Points: data, Lines: fits using equation (\ref{ramsey})}%
\label{ramseyPlot}%
\end{figure}

For the most precise measurement of the resonance frequency we use Ramsey's method of separated oscillatory fields \cite{Ramsey1950}. This method is less sensitive to the field inhomogeneities noted above, and it reduces the resonance linewidth by approximately 40\% compared with a single-pulse measurement of the same over-all duration. The pulse sequence consists of two short $\pi/2$ pulses of duration $\tau$ and angular frequency $\omega$ separated by a period of free evolution $T$. The first $\pi/2$ pulse is applied when the molecular pulse is at antinode $m_1$ and creates a superposition of the two $\Lambda$-doublet components. The coherence evolves freely for a time $T$ at the transition angular frequency $\omega_0$ and develops a phase difference $\delta T$ relative to the microwave oscillator, where $\delta=\omega-\omega_0$. A second $\pi/2$ pulse completes the population transfer with a probability of
\begin{align}
P_{\text{2-pulse}}\left(\delta\right)=&\,\frac{4\pi^2\sin^2\left(\frac{X}{4}\right)}{X^4}\times \notag \\
&\,\left[X\cos\left(\frac{X}{4}\right)\cos\left(\frac{\delta T+\beta}{2}\right)-2\delta\tau\sin\left(\frac{X}{4}\right)\sin\left(\frac{\delta T+\beta}{2}\right)\right]^2\, ,
\label{ramsey}
\end{align}
where $X=\sqrt{\pi^2+4\delta^2\tau^2}$ and $\beta$ is any change in the phase of the microwave field between one pulse and the next (here is no such phase shift if the field is a perfect standing wave). We set $\tau=\SI{15}{\micro\second}$, and choose the free evolution time $T=m\lambda/(2v_0)-\tau$, where $m$ is an integer, so that the molecules travel an integer number of half wavelengths between the start of one pulse and the start of the next, making $\beta=0$ (modulo $\pi$) even for a travelling wave. Figure \ref{ramseyPlot} shows data taken this way to measure the $(1/2^-,1)-(1/2^+,1)$ and $(3/2^-,2)-(3/2^+,1)$ transition frequencies. The figure shows data for several different free evolution times $T$. The lines are fits using the model $b+a P_{\text{2-pulse}}(\delta)$ with $\beta=0$, and with $\tau$ and $T$ set to the values used in the experiment. This leaves only the offset $b$, the amplitude $a$ (negative if $m$ is odd) and the resonance angular frequency $\omega_0$ as fitting parameters. The frequencies measured for different values of $m$ are all in agreement.
\section{Velocity-dependent frequency shifts}\label{Sec:DopplerShifts}

If $\beta$ is not zero, there will be a systematic frequency shift, $\delta f=\beta/\left(2\pi T\right)$. An obvious contribution to $\beta$ comes from the amplitude imbalance between the co-propagating and counter-propagating waves, $\beta=\phi(z_2)-\phi(z_1)$, where $\phi(z)$ is given by equation (\ref{imbal}) and $z_{1,2}$ are the positions of the molecules at the start of the first and second pulses. As mentioned above, we aim to place $z_{1}$ and $z_{2}$ at antinodes of the standing wave. This can be achieved with high accuracy using the field map that we make for each frequency measurement, such as the one shown in figure \ref{fieldmap}, but there will always be some error, and there is a spread in the positions of the molecules. In fact, this spread in positions can usefully be used to map out the position dependence of the phase. Consider a molecule whose arrival time at the detector is $t_d$. It is at position $z_{1}=v t_0$ at the time $t_0$ when the first pulse starts, and at position $z_{2}=v(t_0+\tau + T)$ when the second pulse starts, where $v=D/t_d$ is its speed. For this molecule, the expected systematic shift is
\begin{equation}
\delta f = \frac{1}{2\pi T}\left\{\tan^{-1}\left[\Delta\tan\left(k v(t_0+\tau + T)- k z_0\right)\right] - \tan^{-1}\left[\Delta\tan\left(k v t_0- k z_0\right)\right] \right\}.
\label{fullDopp}
\end{equation}

\begin{figure}%
\centering%
\includegraphics[width=10cm]{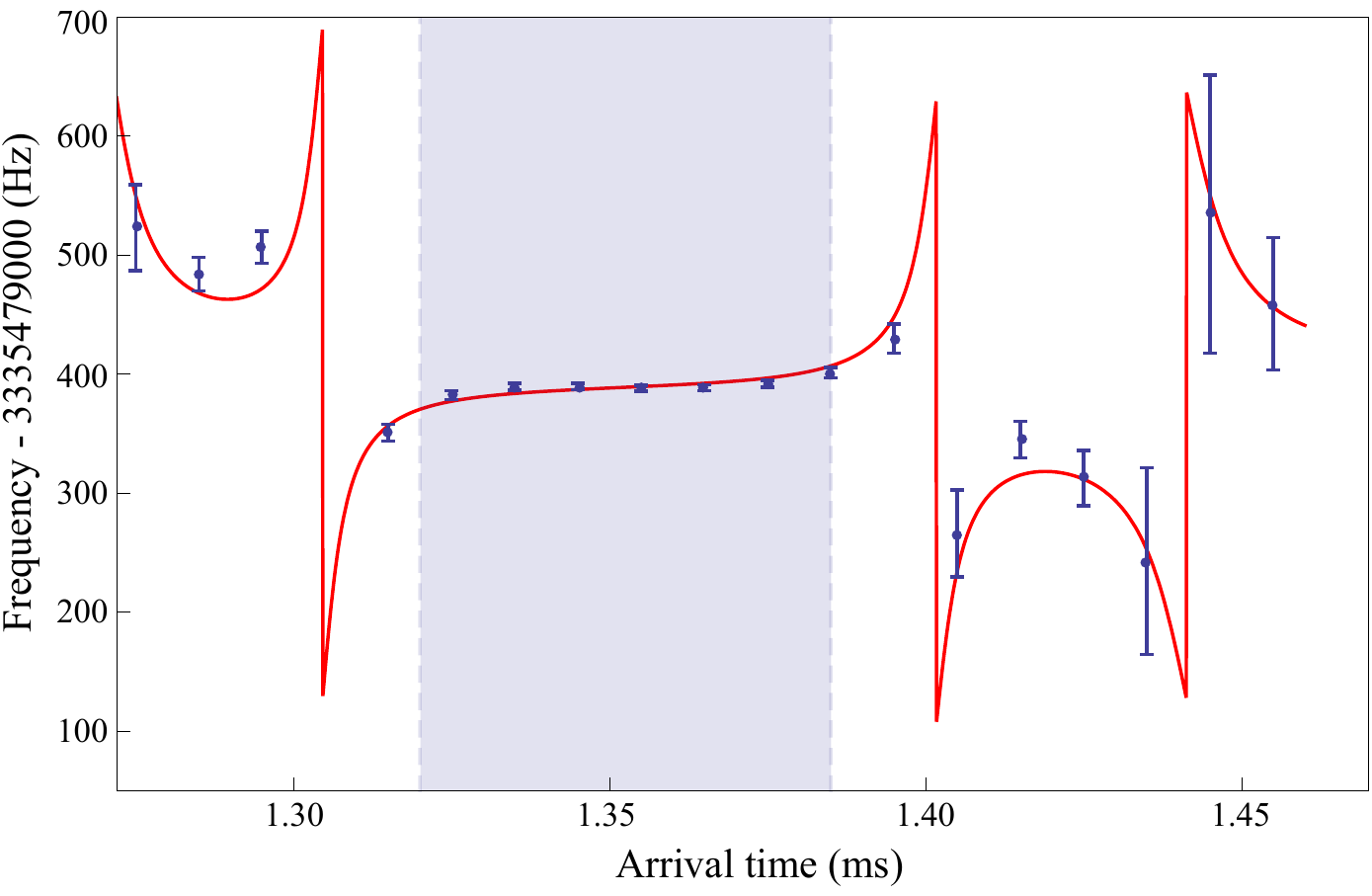}%
\caption{Systematic frequency shift of the $(1/2^+,1)-(1/2^-,1)$ transition as a function of the arrival time of the molecules. For each \SI{5}{\micro\second} interval of the time-of-flight profile, the resonance frequency is measured by fitting equation~(\ref{ramsey}) to Ramsey data. The red solid line is a fit to equation~(\ref{fullDopp}) where the fitting parameters are $z_0$, $\Delta$, and an overall frequency offset. All other parameters are fixed to the values they have in the experiment. The shaded regions indicates the range of arrival times used for the frequency measurements.}%
\label{velModelPlot}%
\end{figure}

We divide the time-of-flight profile into slices \SI{5}{\micro\second} wide. For each slice we find the resonance frequency using the Ramsey method, and plot this against the arrival time. Figure \ref{velModelPlot} shows the data obtained this way for the $(1/2^+,1)-(1/2^-,1)$ transition. We see that the frequency shift is small for molecules near the central arrival time ($\approx 1.35$\,ms) because, when the pulses were applied, these molecules were close to the antinodes where the phase changes very slowly. The molecules that arrive later are moving more slowly, and they are closer to the nodes when the pulses are applied. Molecules arriving at \SI{1.4}{\milli\second} are near the node when the second pulse is applied, and here we see a sudden change in the frequency shift. Those arriving even later, at about \SI{1.44}{\milli\second}, are at a node when the first pulse is applied, and we see another sudden change in the frequency shift. The line in figure \ref{velModelPlot} is a fit to the model $f_0 + \delta f$ where $\delta f$ is given by equation (\ref{fullDopp}) and $f_{0}$ is the resonance frequency for molecules in the centre of the pulse. We fix $t_0$, $\tau$, $T$ and $D$ to the values used in the experiment, while $f_0$, $z_0$ and $\Delta$ are fitting parameters. From the fit, we find $\Delta =-0.079\pm 0.002$. This model agrees remarkably well with the data, showing that we have excellent control over position-dependent phases in the experiment. For frequency measurements, we use only those molecules that arrive during the period indicated by the shaded region in figure \ref{velModelPlot}, where the frequency changes by less than 20\,Hz. We note that for the $J=3/2$ measurements, a plot similar to figure \ref{velModelPlot} showed no frequency dependence at the 20\,Hz level over the entire range of arrival times.

For molecules near the centre of the pulse, we can find a simple expression for the systematic frequency shift discussed above. Let $z_1=z_0+\delta z_0$ and $z_2=z_1 + (1+\epsilon)m\lambda/2$, where $z_0$ is now the position of any antinode, while $\delta z_0/\lambda \ll 1$ and $\epsilon \ll 1$ account for the imperfect timing of the microwave pulses due to the uncertainty in the velocity of the molecule. Expanding the trigonometric functions in equation (\ref{fullDopp}) to first order in these small quantities, we find that the systematic frequency shift is
\begin{equation}
\delta f \simeq\epsilon\,\Delta\left(\frac{v}{\lambda}\right).
\label{imbalEa}
\end{equation}

This result shows that the frequency shift is independent of $\delta z_0$ to first order and that the first-order Doppler shift ($v/\lambda$) is suppressed by the product of two small quantities - $\Delta$, which is the imbalance factor, and $\epsilon$ which is the fractional error in setting the interaction length to an integer number of half-wavelengths. We can set upper limits to $\Delta$ in three ways: the finesse of the transmission line resonator (figure \ref{stand}), the power needed to drive $\pi$-pulses for each of the resolved Doppler-shifted components (figure \ref{sPlot}), and fitting to data such as that in figure \ref{velModelPlot}. The first method gives $\Delta < 0.09$ but does not give a tight constraint because the finesse of the resonator is only partly determined by the reflection at the open end, the other part being the transmission at the input end. The other two methods give consistent results as follows: $\Delta < 0.08$ for the $J=1/2$ components at 3335 and \SI{3349}{\mega\hertz}, $\Delta < 0.005$ for the $J=1/2$ component at \SI{3264}{\mega\hertz}\footnote{After measuring the first two frequencies the plates were cut to a new length to measure the \SI{3264}{\mega\hertz} line. This must be the reason for the change in $\Delta$.}, and $\Delta < 0.08$ for all the $J=3/2$ components. An upper limit to $\epsilon$ comes from the maximum fractional uncertainty in determining the central speed of the pulse, which we estimate to be 0.03. In addition, molecules with different speeds have different values of $\epsilon$. We select molecules from the time-of-flight profile with arrival times in the range $t_0\pm\delta t$, where $t_0$ is the most probable arrival time and $\delta t\simeq 0.02 t_0$, and so the spread in $\epsilon$ values is $\pm 0.02$. Thus, the maximum possible value of $\epsilon$ in the experiment is 0.05. For the $J=1/2$ measurement the upper limit to this systematic shift is \SI{0.04}{\hertz\per(\meter\per\second)}. Note that for $J=3/2$, the shift is about 5 times smaller for the same $\epsilon$ and $\delta$, because of the longer wavelength.

There is also a second contribution to the velocity-dependent frequency shift which stems from the motion of the molecules during the two short $\pi/2$ pulses. Consider first the interaction with a \emph{travelling} wave tuned to the resonant angular frequency $\omega_0$. For a stationary molecule, the phase difference between the microwave oscillator and the oscillating dipole is $\pi/2$, but for a moving molecule there is an additional contribution to this phase difference due to the Doppler shift $\delta_D=2\pi v/\lambda$. To second order in $\delta_D$, this phase difference is $[1-\tan(\Omega\tau/2)/(\Omega\tau)]\delta_D\tau$. Due to this phase shift, the population transfer is maximized by slightly detuning the microwave oscillator. To find the resulting systematic frequency shift, we include the Doppler shift in the expression for the Ramsey lineshape, expand this to second order in both the detuning and the phase shift, $\beta$, between the two pulses, and then find the value of $\beta$ that maximizes the population transfer. When $T\gg\tau$ the frequency shift is

\begin{equation}
\delta f=\left(1+\frac{2(\tan(\Omega\tau)-\csc(\Omega\tau))}{\Omega\tau}\right)\frac{v_0}{\lambda}\frac{\tau}{T}.
\label{doppR}
\end{equation}
For our case, where $\Omega\tau=\pi/2$, the bracketed quantity is $(1-4/\pi)$. We see that the Doppler shift is suppressed by the small quantity $\tau/T$, which we may have expected since the microwave field is only applied for this fraction of the time. When there are two counter-propagating waves, we might expect the above expression to be further suppressed by the imbalance factor $\Delta$, and our numerical modelling shows that this is indeed the case. With $\Delta<0.08$ and typical values of $\tau$ and $T$, we find the shift to be less than \SI{0.01}{\hertz\per(\meter\per\second)}.

\begin{figure}%
\centering%
\includegraphics[width=\columnwidth]{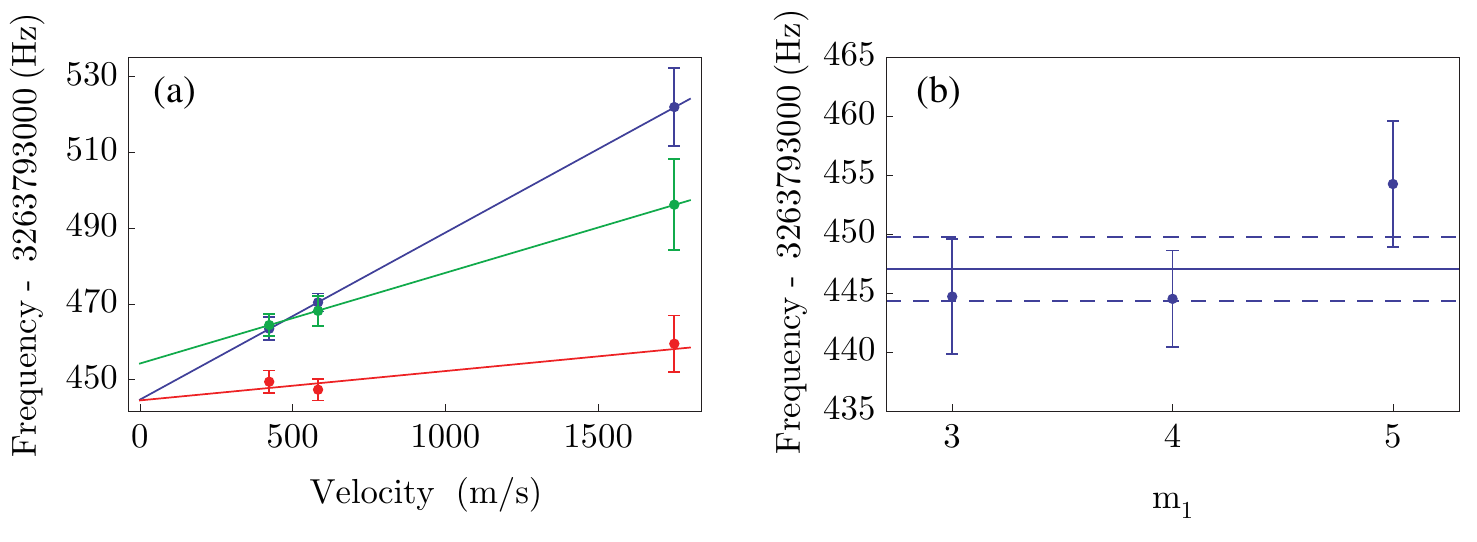}%
\caption{(a) Resonance frequency of the $(1/2^+,1)-(1/2^-,0)$ transition as a function of the most probable velocity $v_0$, for $m_1=3,4,5$ (blue, red, green). (b) Extrapolated zero-velocity frequencies for the three values of $m_1$. They agree within the uncertainty of the linear fits and we take the weighted mean (solid line). The 1-$\sigma$ standard error of the weighted mean is shown by the dashed lines.}%
\label{velSys}%
\end{figure}

There are other possible velocity-dependent frequency shifts in addition to those discussed above. For example, a position-dependence of the polarization can produce a frequency shift proportional to the velocity. To control these shifts, we measure the transition frequency for at least three different velocities. Figure~\ref{velSys}(a) shows the measured frequency of the $(1/2^+,1)-(1/2^-,0)$ transition as a function of $v_0$ for three different values of $m_1$ (the antinode used for the first microwave pulse). For each data point in the figure, we average together at least three measurements with different values of $m$, since we find no dependence on $m$. We see that the measured frequency depends linearly on the velocity of the molecules and that the gradient $df/dv_0$ differs for different values of $m_1$. The largest gradient observed is $0.05\pm 0.01$\si{\hertz\per(\meter\per\second)}. After extrapolating to zero velocity, the measurements using various $m_1$ are all in agreement, as shown in figure~\ref{velSys}(b). We average together these zero-velocity results to obtain the final transition frequency, and we do this for all seven frequencies measured. For the $J=3/2$ measurements, the largest velocity-dependence we observed was $0.03\pm 0.01$\si{\hertz\per(\meter\per\second)}, and we observed no dependence on $m_1$.

\section{Zeeman shifts}

Next, we consider systematic frequency shifts due to magnetic fields in the interaction region. For small magnetic fields, the Zeeman splitting is linear and symmetric about the line centre. The amplitudes of the components are also symmetric if the microwave field is linearly polarized. Frequency shifts arise due to circular polarization components and/or higher-order Zeeman shifts. To minimize these shifts the interaction region is magnetically shielded (see figure \ref{ms32}).

In the $J=1/2$ state the magnetic moments arising from the orbital and spin angular momenta are very nearly equal and opposite and the $g$-factor is of order $10^{-3}$. For the $J=1/2$ measurements, we used just a single layer mu-metal shield to reduce the background magnetic field to acceptable levels. We applied fields as large as \SI{50}{\micro\tesla} in each direction and observed no shift of the frequencies measured using the Ramsey method, at the \SI{1}{\hertz} level. Since the residual magnetic field is far smaller than this, Zeeman shifts are negligible in the $J=1/2$ measurements.

\begin{figure}%
\centering
\includegraphics[width=\columnwidth]{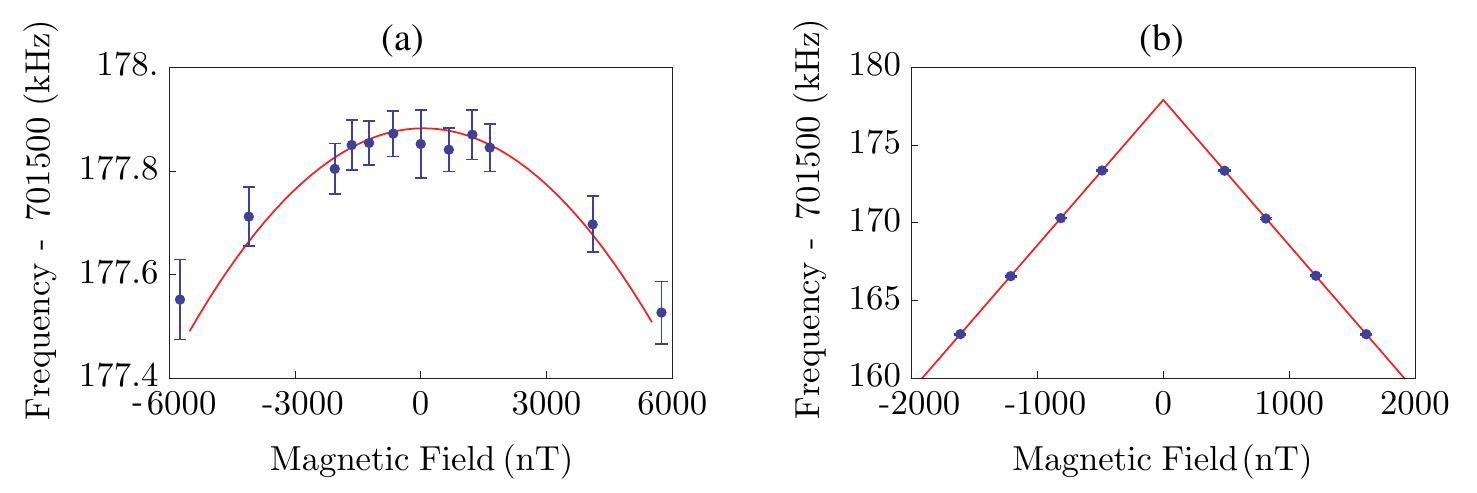}%
\caption{Zeeman shifts of the $(3/2^+,2)-(3/2^-,2)$ transition, measured using single microwave pulses of~\SI{780}{\micro\second} duration. (a) Quadratic Zeeman shift of the $\Delta M_F=0$ transitions versus field applied along $y$. Line: fit to a quadratic model. (b) Linear Zeeman shift of the $\Delta M_F=\pm1$ transitions versus field along $x$. We plot the component that shifts to lower frequency, which is the $\Delta M_F=+1 (-1)$ component for negative (positive) values of the field. Line: fit to a linear model.}%
\label{twoB}%
\end{figure}

The g-factor is much larger in the $J=3/2$ state - $g=1.081$ for $F=1$ and $g=0.648$ for $F=2$ - and so for the $J=3/2$ measurements we improved the shielding by using a two-layer shield. Inside the shields we create homogeneous magnetic fields along $x$ and $y$ using Helmholtz coils and along $z$ using a solenoid. These coils are calibrated with a fluxgate magnetometer. To measure the Zeeman splitting of the hyperfine components, we apply large enough magnetic fields to resolve the splitting of the spectral line obtained using a single microwave pulse of \SI{780}{\micro\second} duration. Figure \ref{twoB}(a) shows the Zeeman shift of the $(3/2^+,2)-(3/2^-,2)$ hyperfine component as a function of the magnetic field applied in the $y$-direction, parallel to the polarization of the microwaves. In this case, only the $\Delta M_F=0$ components are driven, and the shift is quadratic and negative. This shift is due to mixing of the $M_F=0,\pm 1$ levels of $F=2$ with those same components in $F=1$. In the f state, $F=1$ lies lower (see figure \ref{chlevel}) and the mixing raises the energy of $F=2$. The opposite is true for the e state, where the shift is also far smaller because of the larger hyperfine splitting. Therefore, the shift of this transition is negative and is determined mainly by the shift of the lower $F=2$ level. We measure a quadratic Zeeman shift of $-12.6\pm 1.1$~Hz/($\mu$T)$^2$, which is consistent with our calculation. Figure \ref{twoB}(b) shows the Zeeman shift of the same hyperfine component as a function of the magnetic field applied in the $x$-direction. In this case, we drive $\Delta M_F=\pm1$ transitions and observe a linear Zeeman shift of $9.34\pm0.28$~Hz/nT, again consistent with our calculation. Here, the error is dominated by the uncertainty in the calibration of the magnetic field coils. We determine the residual magnetic field averaged over the interaction region by measuring the change in Zeeman shift upon reversal of the applied magnetic fields. We also look for any broadening of the single-pulse lineshape due to residual Zeeman splittings. Together, these set upper limits to the background magnetic field of 3, 56, and \SI{25}{\nano\tesla} along $x$, $y$ and $z$ respectively.

In the frequency measurements made using the Ramsey method, we look for a shift in the resonance frequency as a function of applied magnetic fields. For the $(3/2^+,2)-(3/2^-,2)$ transition we measure a maximum gradient of $-0.017\pm0.013$~Hz/nT for fields applied along $x$, $-0.036\pm 0.005$~Hz/nT along $y$ and $-0.005\pm0.003$~Hz/nT along $z$. Using these, and the upper limits for background magnetic fields, we get upper limits for systematic frequency shifts of \SI{0.1}{\hertz}, \SI{2}{\hertz} and \SI{0.2}{\hertz} for residual fields along $x$, $y$ and $z$ respectively. Adding these in quadrature gives a total systematic uncertainty due to uncontrolled magnetic fields of \SI{2}{\hertz}. We assume the same systematic uncertainty for the $(3/2^+,1)-(3/2^-,1)$ transition due to the similar Zeeman structure.\\ \indent
For the $(3/2^+,1)-(3/2^-,2)$ transition we could not rule out gradients as large as 0.1~Hz/nT along x, 0.2~Hz/nT along y and 0.05~Hz/nT along z. Multiplying these by the upper limits to the residual field gives a systematic uncertainty of \SI{11}{\hertz}. We assume the same uncertainty for the $(3/2^+,2)-(3/2^-,1)$ transition due to its similar Zeeman structure.

\section{Stark shifts}

\begin{figure}%
\centering
\includegraphics{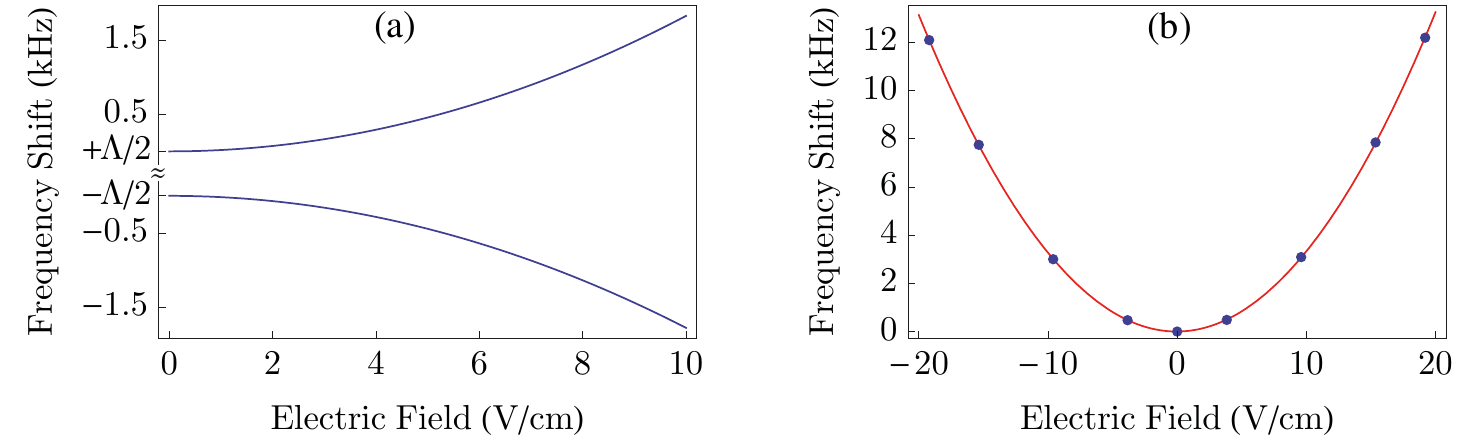}%
\caption{(a) Calculated Stark shift of the $\Lambda$-doublet states for low electric fields. There is no dependence on $F$ or $M_F$. (b) Measured frequency shift of the $(1/2^+,1)-(1/2^-,1)$ transition.}%
\label{stark1}%
\end{figure}

Figure \ref{stark1}(a) shows the calculated Stark shift of the two $J=1/2$ $\Lambda$-doublet levels in low electric fields. The two levels shift oppositely and quadratically. Using a dipole moment of 1.46\,D \cite{Phelps1966} we calculate a shift of $\mp 17.8$~Hz/(V/cm)$^2$. For $J=1/2$ the shift has virtually no dependence on $F$ or $M_F$. To measure the Stark shift of the microwave transition we use single, long microwave pulses, apply a DC voltage to one of the plates of the transmission line via a biased tee, and record the transition frequency as a function of the electric field. Figure \ref{stark1}(b) shows our data for the $(1/2^+,1)-(1/2^-,1)$ transition. The line is a fit to a parabola, $\delta f = a (E-E_b)^{2}$, where $E$ is the applied field and $E_b$ the background electric field. This fit gives a Stark shift of $33 \pm 3$~Hz/(V/cm)$^2$, where the error is dominated by the systematic uncertainty in the plate spacing. The background field is consistent with zero, and from the uncertainty in this field we obtain an upper limit to uncontrolled Stark shifts of \SI{0.1}{\hertz}. For $J=3/2$, the Stark shift depends on $F$ and $M_F$. Using the same procedure as before we obtain a systematic uncertainty of \SI{0.2}{\hertz} for the $J=3/2$ measurements.

\section{Other systematic uncertainties}

We test for systematic frequency shifts that depend on the microwave power by using shorter $\pi/2$ pulses in a Ramsey experiment. We reduce the pulse length from 15 to \SI{4}{\micro\second}, increasing the power by a factor of 14, and do not find any frequency shift. We also do not find any dependence on the probe laser detuning.

Scanning the microwave frequency can change the power in the resonator and this can lead to a systematic frequency shift. Consider a worst-case model where the molecular signal depends linearly on the microwave power and the resonator is badly tuned so that the molecular frequency is half way down the side of a resonance, where the gradient is steepest. Suppose the molecular signal is a Gaussian with a standard deviation $w$, and the transmission line resonances are Lorentzian with FWHM $W$. We find that there is a systematic shift of $|\Delta f|=2w^2/W$. In the experiment, typical values are $w\simeq 1.5$\,kHz and $W \simeq 30$\,MHz, leading to a shift of only \SI{0.15}{\hertz}. In reality, the shift is much smaller for two reasons. First, we tune the transmission line to be on resonance at the molecular frequency. Second, we choose the power that maximizes the molecular signal, i.e. a $\pi$-pulse in a single-pulse measurement or $\pi/2$-pulses in a Ramsey measurement, and so the signal has no first derivative with respect to power.

Unwanted frequency sidebands may be produced, for example by the microwave oscillator or by the switching electronics, and these can lead to systematic shifts if there is an asymmetry in the amplitudes of the sidebands. We have measured the frequency spectrum and find no sidebands down to -40\,dB within \SI{1}{\mega\hertz} of the oscillator frequency. For our parameters, we estimate that the systematic frequency shift will be largest for an asymmetric sideband with an offset of \SI{40}{\kilo\hertz} \cite{Audoin1978}. Then, if the amplitude of this single sideband is -40\,dB, the resulting frequency shift is only \SI{10}{\micro\hertz}.

Systematic frequency shifts due to blackbody radiation, the motional Stark effect, the second-order Doppler shift, and collisional shifts, are also all negligible at the current accuracy level.

\section{Conclusions}

Table \ref{finalValues} gives our final transition frequencies, reproduced from \cite{Truppe2013}. We add the statistical and systematic uncertainties in quadrature to give the total uncertainty.

\begin{table}
\centering
\begin{tabular}{c|c}
\hline\hline
Transition & Frequency (\si{\hertz})\\
\hline
$(1/2^+,1)-(1/2^-,1)$ & $3335479356\pm 3$\\
$(1/2^+,0)-(1/2^-,1)$ & $3349192556\pm 3$\\
$(1/2^+,1)-(1/2^-,0)$ & $3263793447\pm 3$\\
$(3/2^+,2)-(3/2^-,2)$ & $701677682\pm 6$\\
$(3/2^+,1)-(3/2^-,1)$ & $724788315\pm 16$\\
$(3/2^+,1)-(3/2^-,2)$ & $703978340\pm 21$\\
$(3/2^+,2)-(3/2^-,1)$ & $722487624\pm 16$
\end{tabular}
\caption{The measured $\Lambda$-doublet transition frequencies with their $1\sigma$ uncertainties.}
\label{finalValues}
\end{table}

Our method of microwave spectroscopy is exceptionally versatile and accurate. The method can be used for any molecule that can be produced in a pulsed supersonic beam, and the same apparatus can be used over a very wide frequency range, including low frequencies where a conventional microwave cavity would be too large. Ramsey spectroscopy is more commonly done using two separate cavities with their axes perpendicular to the molecular beam direction. The phase difference between the two cavities then needs to be accurately controlled, which can be difficult to achieve. In the transmission line resonator, control over the relative phase of the two pulses is straightforward because the field is supported by a single structure. With high-Q cavities it is necessary to scan the cavity in synchronism with the microwave frequency. By contrast, the transmission line resonances are broad enough that it is not necessary to tune the line as the frequency is scanned. We emphasize the importance of choosing the plate width and spacing so that higher-order modes, include the `leaky modes', are strongly attenuated. With only the TEM mode able to propagate, the field is very well controlled. In our setup, molecular resonance frequencies can be measured either using a single, long microwave pulse, or using the Ramsey method of two short pulses separated in time. These pulses can be applied when the molecules are at any position along the transmission line and so the molecules can be used to map out the amplitude and phase of the microwave field, providing an exceptional degree of control. For this mapping method to work well it is important to produce short, cold pulses of molecules, and to use a detector with adequate time resolution. Our CH source produces pulses that are just a few millimetres in length and with a translational temperature as low as 400\,mK.

Because the field is a standing wave, Doppler shifts are very strongly suppressed in the experiment. As described in section \ref{Sec:DopplerShifts}, the residual Doppler shift in the Ramsey measurements is proportional to two small factors, one being the imbalance in amplitude between the two counter-propagating waves, the other being the fractional error in setting the interaction length to an integer number of half-wavelengths. In this experiment, we observed velocity-dependent shifts at the level of \SI{0.05}{\hertz\per(\meter\per\second)} or less. We measured and eliminated these shifts by varying the beam velocity. The experiment reached a precision of \SI{3}{\hertz}, limited mainly by the statistical uncertainty in extrapolating to zero velocity. The velocity-dependent shifts could be reduced by improving the reflection at the end of the transmission line to reduce the amplitude imbalance between the counter-propagating waves, and by improving the way the microwaves are launched into the transmission line to eliminate field non-uniformities in this region. A Stark decelerator \cite{Bethlem1999} could also be used to reduce the velocity of the beam by a factor of 10, giving both longer interaction times and improved velocity control \cite{Hudson2006}.

Individual frequency measurements reached a statistical uncertainty of \SI{1}{\hertz} within about 1 hour. This was partly limited by a background of scattered laser light that reaches the detector, and partly by shot-to-shot fluctuations of the source. The background could be reduced by improving the shape of the probe laser mode, and the source noise could be reduced by using a second laser-induced-fluorescence detector upstream of the experiment to record the number of molecules produced in each shot. The photon shot noise limit could then be reached, giving an uncertainty of \SI{1}{\hertz} in just a few minutes of integration.

 \section*{Acknowledgements}
We thank Ben Sauer, Jony Hudson and Heather Lewandowski for their help and advice. We are indebted to Jon Dyne, Steve Maine and Valerijus Gerulis for their expert technical assistance. This work was supported by the EPSRC and the Royal Society.

\newpage
\bibliographystyle{elsarticle-num}
\bibliography{library1.bib}
\end{document}